\documentclass[12pt]{article}
\usepackage[utf8]{inputenc}
\usepackage{indentfirst}
\usepackage{amsmath,amsthm,amssymb}
\usepackage{graphicx}
\usepackage{setspace}
\usepackage{hyperref}
\usepackage[all]{hypcap}
\usepackage[affil-sl,auth-sc]{authblk}
\pdfoutput=1 
\title{The domain wall partition function for the Izergin--Korepin 19-vertex model 
at a root of unity}
\author{A Garbali%
\thanks{Electronic address: alexandr.garbali@unimelb.edu.au}}
\affil{Laboratoire de Physique Th\'eorique et Hautes \'Energies, CNRS 
UMR 7589 and Universit\'e Pierre et Marie Curie (Paris 6), 4 place 
Jussieu, 75252 Paris cedex 05, France}
\date{}
\begin{document}

\maketitle

\begin{abstract}
We study the domain wall partition function $Z_N$ for the $U_q(A_2^{(2)})$ 
(Izergin--Korepin) 
integrable $19$-vertex model on a square lattice of size $N$. $Z_N$ is a symmetric function 
of two sets of parameters: horizontal $\zeta_1,..,\zeta_N$ and vertical $z_1,..,z_N$
rapidities. For generic values of the parameter $q$ we derive the recurrence relation for 
the domain wall partition function relating $Z_{N+1}$ to $P_N Z_N$, where $P_N$ is the 
proportionality factor in the recurrence, which is a polynomial symmetric in two sets of 
variables $\zeta_1,..,\zeta_N$ and $z_1,..,z_N$. After setting $q=e^{i\pi/3}$ 
the recurrence relation simplifies and we solve it in terms of a Jacobi--Trudi-like determinant 
of polynomials generated by $P_N$. 
\end{abstract}

\section{Introduction}\label{secti}
We are studying here a particular object of the nineteen vertex model of the $U_q(A_2^{(2)})$ 
quantum group, also called the Izergin--Korepin model. The object of our interest is 
the partition function of the model on a square lattice in a $N\times N$ square region with 
domain wall boundary conditions. 

Our work is motivated by the domain wall partition function (DWPF) for the six vertex model 
$Z_{6v}$, constructed using the $R$-matrix of the $U_q(A^{(1)}_1)$ quantum group. 
This partition function satisfies a set of recurrence relations found by Korepin \cite{Korepin}. These recurrence relations were solved by Izergin \cite{Izergin}. The solution is written in a form of a determinant which is called the Izergin--Korepin (IK) determinant. 
In statistical physics the six vertex model represents a model for two dimensional ice, 
which shows interesting critical phenomena (see \cite{Baxter}).
The partition function $Z_{6v}$ plays a very important role in the field of integrable 
models. It is a crucial object in the theory of correlation functions for integrable 
spin chains \cite{KBI} such as the $XXZ$ spin-$1/2$ chain (see also \cite{KMT}). 
In combinatorics it allowed one to count the alternating sign matrices \cite{Kuperberg1,Kuperberg2}. To compute the domain wall partition functions for other vertex 
model is a very complicated problem. One of the main results generalizing the six vertex domain wall partition 
function (DWPF) is due to \cite{CFK}, where the $U_q(A^{(1)}_1)$ higher spin generalization of the DWPF is obtained in 
a determinant form.

Inspired by these and other results we address the question 
of computing the domain wall partition function for the $U_q(A^{(2)}_2)$ 
nineteen vertex model.
This model is an integrable model associated to the Dodd--Bullough--Mikhailov equation,
also known as Jiber--Mikhailov--Shabat model \cite{DB,ZS,Mik}. Izergin and Korepin computed the classical 
and quantum $R$-matrices for this model \cite{IK}. 
The quantum $R$-matrix defines the Izergin--Korepin vertex model. 
The IK\footnote{The abbreviation IK can be a bit misleading here. Both the model and the 
object that we want to compute contain the IK part in their short names.} 
$R$-matrix has nineteen non zero entries, which correspond to the
nineteen possible vertex configurations (see Fig. \ref{vertices}).  
We use this $R$-matrix to build $N$ by $N$ lattice configurations which have the domain wall boundary conditions Fig. \ref{DWPF}. 
The sum of all such configurations we call the domain wall partition function $Z_N$. In 
order to compute $Z_N$ we use the ideas from the six vertex model. First, we establish the 
recurrence relation for the partition function and then try to find its unique solution. 
In the case when the deformation parameter $q$ is generic we cannot find a compact expression for $Z_N$. 
However, when $q=e^{i\pi/3}$ we are able to find a determinant expression.

In Section \ref{sect2} we shortly discuss the DWPF for the six vertex model. In 
Section \ref{sect3} we move to the IK model. In Section \ref{sect4} we derive the 
recurrence relation using the vanishing properties of the weights of the $R$-matrix. 
The solution 
to this recurrence relation at $q=e^{i\pi/3}$ is presented in Section \ref{sect5}. The proof follows in Section \ref{sect6} and 
with a conclusion we finish.   

\section{Six-vertex model with domain wall boundary}\label{sect2}
For the computation of the IK determinant for the six vertex model check the papers 
\cite{Izergin,Korepin}. 
Here we present a short discussion for convenience.

\begin{figure}[htb]
\centering
\includegraphics[width=0.6\textwidth]{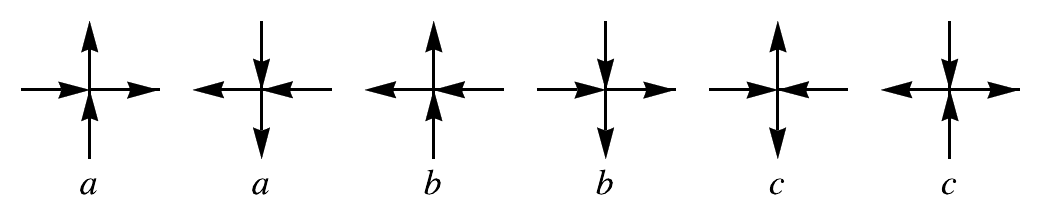}
\caption{The six vertices of the six vertex model. The letters $a,~b$, and $c$ are the 
weights of the corresponding vertices.}
\label{6vconf}
\end{figure}

\begin{figure}[htb]
\centering
\includegraphics[width=0.3\textwidth]{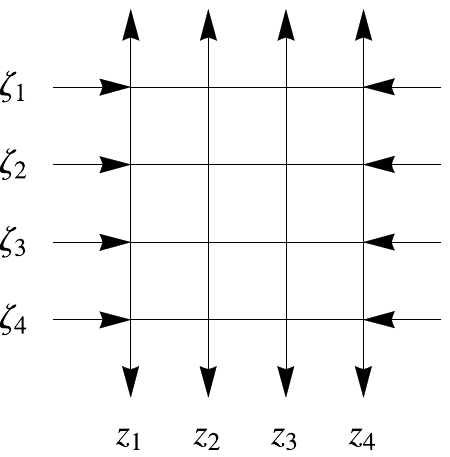}
\caption{The domain wall boundary conditions. The parameters $\zeta_1,..,\zeta_4$ are associated 
to the horizontal lines, while the parameters $z_1,..,z_4$ are associated to the vertical lines.}
\label{DWPF}
\end{figure}

\begin{figure}[htb]
\centering
\includegraphics[width=0.4\textwidth]{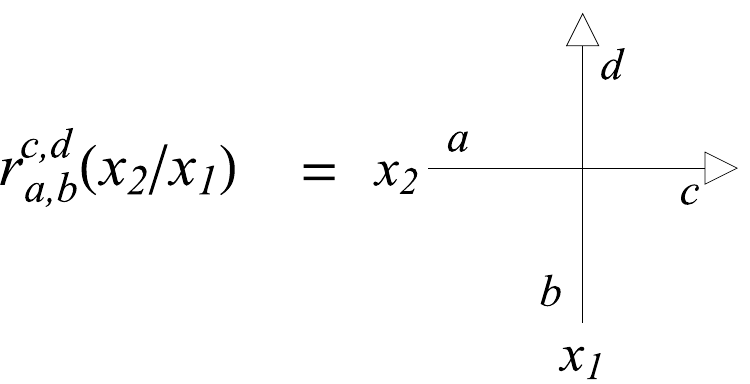}
\caption{Components $r_{a,b}^{c,d}(x_2/x_1)$. The blank arrows here define the orientations of the vector spaces. 
These arrows should not be confused with the arrows in Fig. \ref{6vconf} which are used to denote the configurations of edges.}
\label{fRc}
\end{figure}

\begin{figure}[htb]
\centering
\includegraphics[width=0.4\textwidth]{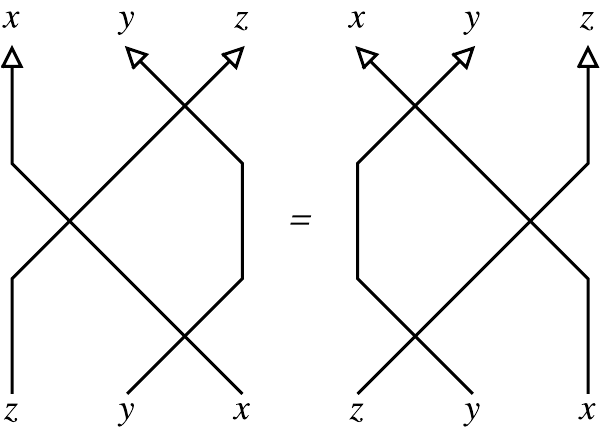}
\caption{The Yang--Baxter equation. The spectral parameters $x,~y$ and $z$ are carried 
by the corresponding vector spaces.}
\label{YB}
\end{figure}

The problem is to sum up configurations which are built by 
choosing for each vertex of a square $N\times N$ lattice one of the six vertices from 
Fig. \ref{6vconf}. 
A configuration thus constructed will have on each edge one of the two states: a left arrow or 
a right arrow if the edge is horizontal and an up arrow or a down arrow for a vertical edge. 
We then impose the domain wall boundary conditions which are depicted in Fig. \ref{DWPF}.  
Each vertex on this lattice has a position $(i,j)$ where horizontal position $i$ is counted 
rightwards, and the vertical position $j$ is counted downwards starting from the top left corner. 
The weight of the vertex at position $(i,j)$ is denoted by $w_{i,j}$ and takes one of the three 
values $a_{i,j},~b_{i,j}$ or $c_{i,j}$. The weight of a configuration $\varepsilon$ on a square 
domain of size $N\times N$ will be the product of all weights of its vertices
\begin{align}
\prod_{1\leq i,j\leq L}w^{(\varepsilon)}_{i,j}. \nonumber
\end{align}
The IK partition function is the sum over all configurations (states) $\varepsilon$: 
\begin{align}
Z_{6v}=\sum_{\varepsilon\in \text{states}}\prod_{1\leq i,j\leq N}w^{(\varepsilon)}_{i,j}.
\end{align}
The weights $a,~b$ and $c$ are encoded in the $R$-matrix. The $R$-matrix acts on two 
vector spaces labeled by $i$ and $j$, which carry spectral parameters $x_i$ and $x_j$, 
thus we write $R_{i,j}(x_i,x_j)$. We write the $R$-matrix  in the spin basis:
$v_+=(1,0)$ and $v_-=(0,1)$, where $v_+$ corresponds to an up arrow 
if the edge is vertical and a right arrow if the edge is horizontal, similarly $v_{-}$ 
corresponds to a down arrow if the edge is vertical and a left arrow if the edge is horizontal.
\begin{align}\label{Rm}
R_{i,j}(x_i,x_j)=\left(
\begin{array}{cccc}
 a\left(x_i,x_{j}\right) & 0 & 0 & 0 \\
 0 & b\left(x_i,x_{j}\right) & c\left(x_i,x_{j}\right) & 0 \\
 0 & c\left(x_i,x_{j}\right) & b\left(x_i,x_{j}\right) & 0 \\
 0 & 0 & 0 & a\left(x_i,x_{j}\right) \\
\end{array}
\right)
\end{align} 
In fact the integrable $R$-matrix depends on the ratio 
of the spectral parameters: $R_{i,j}(x_j/x_i)= R_{i,j}(x_i,x_j)$. 
Using the matrix units $e_{a,b}$ as a basis for the matrices acting in $\mathbb{C}^2$ we can write
(the indices in the summations in this section take values $-$ and $+$):
\begin{align}\label{Rcomp}
R(x_2/x_1)=\sum_{a,b,c,d} r_{a,b}^{c,d}(x_2/x_1)e_{a,c}\otimes e_{b,d},
\end{align}
where the components of the $R$-matrix are denoted by $r_{a,b}^{c,d}$; furthermore we 
will use their graphical representation Fig. \ref{fRc}. We will also need 
the $\check{R}$-matrix: $\check{R}=P R$, where $P$ is the permutation matrix:
\begin{align}\label{perm}
P=\sum_{a,b}e_{a,b}\otimes e_{b,a},
\end{align}
so we have:
\begin{align}\label{Rccomp}
\check{R}(x_2/x_1)=\sum_{a,b,c,d} \check{r}_{a,b}^{c,d}(x_2/x_1)e_{a,c}\otimes e_{b,d},
\end{align}
\begin{figure}[htb]
\centering
\includegraphics[width=0.4\textwidth]{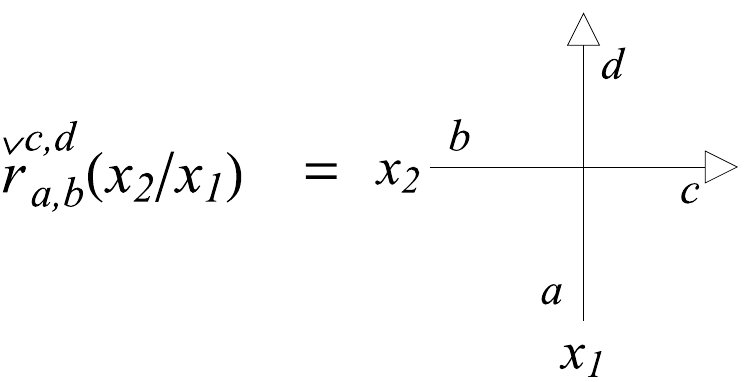}
\caption{Components $\check{r}_{a,b}^{c,d}(x_2/x_1)$.}
\label{fRch}
\end{figure}
Graphically, the components of $\check{R}$ are presented in Fig. \ref{fRch}. 
The integrable $R$-matrix satisfies the Yang--Baxter equation. 
Using the schematic notation of the $\check{R}$-matrix this equation can be drawn as 
in Fig. \ref{YB}.
The Yang--Baxter equation corresponding to Fig. \ref{YB} is written then as
\begin{align}\label{YBm}
\check{R}_{i+1}(y,x)\check{R}_{i}(z,x)\check{R}_{i+1}(z,y)=
\check{R}_{i}(z,y)\check{R}_{i+1}(z,x)\check{R}_{i}(y,x),
\end{align}
where $\check{R}$-matrices here are: $\check{R}_i=\check{R}\otimes Id$ 
and $\check{R}_{i+1}=Id\otimes \check{R}$.
This equation restricts the possible weighs of the vertices. The solution reads:
\begin{align}\label{weights}
a(x_i,x_j )=\frac{q^2 x_i^2-x_j^2}{\left(q^2-1\right) x_i x_j},~~~
b\left(x_i,x_j\right)=\frac{q \left(x_i^2-x_j^2\right)}{\left(q^2-1\right) x_i x_j},~~~  
c\left(x_i,x_j\right)=1.
\end{align}

\begin{figure}[htb]
\centering
\includegraphics[width=1\textwidth]{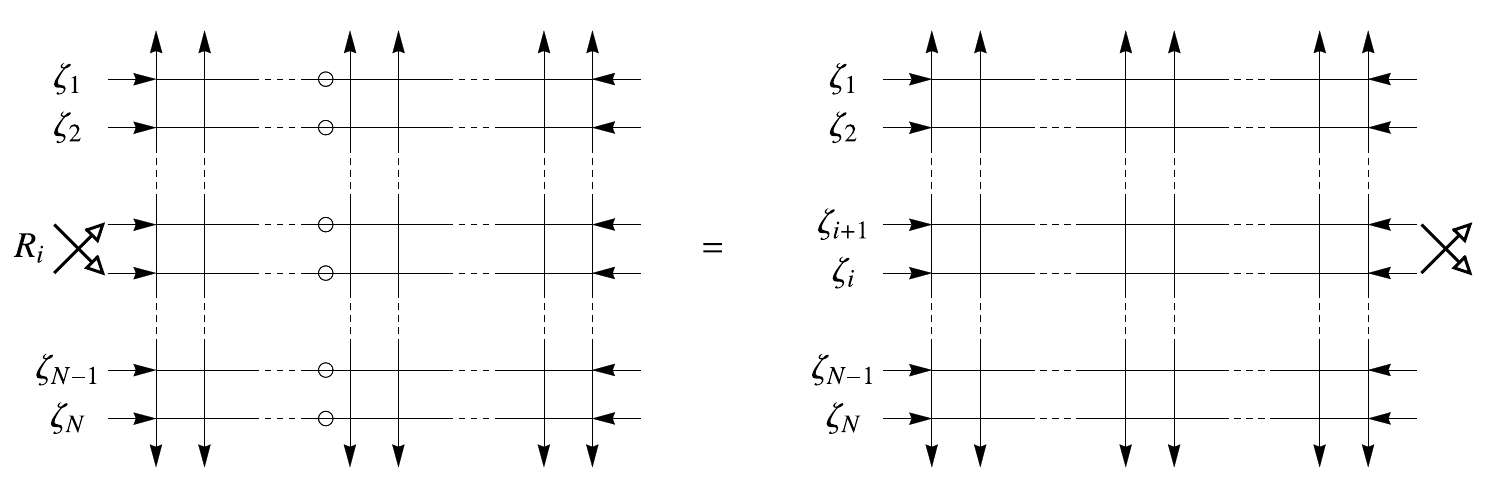}
\caption{On the left side of this equation one must use the Yang--Baxter equation 
Fig. \ref{YB} 
to push through the $\check{R}$-matrix. The boundary conditions are such that on the both sides of this 
equation there is only one term of the $\check{R}$-matrix that contributes, i.e. the vertex with the 
weight $a$. The symmetry in $z$'s is proven similarly.}
\label{Sym}
\end{figure}

In the domain as on Fig. \ref{DWPF} there are $N$ horizontal spaces carrying $N$ parameters 
$\zeta_1,..,\zeta_N$ and $N$ vertical spaces carrying 
$N$ parameters $z_1,..,z_N$. The latter parameters are called inhomogeneities and the model 
therefore is called the inhomogeneous six vertex model. From the form of the weights 
Eq. (\ref{weights}) we see that the partition function $Z_{6v}$ is a polynomial in $z$'s and 
$\zeta$'s divided by a common denominator that we neglect in what follows.
In fact, $Z_{6v}$ is symmetric separately in $z$'s and in $\zeta$'s. It can be seen by applying 
the $\check{R}_{i,i+1}$ matrix to Fig. \ref{DWPF} and using repeatedly the Yang--Baxter equation. 
If the $\check{R}$-matrix is applied at a position $i$ from below or above of the domain 
Fig. \ref{DWPF} this action will switch two rapidities $z_i$ and $z_{i+1}$, if it is applied 
from the sides $\zeta_i$ and $\zeta_{i+1}$ will be switched (see Fig. \ref{Sym}).   

Now we present the computation of the domain wall 
partition function $Z_{6v}$ for the six-vertex model. Before proceeding let us normalize $Z_{6v}$ as follows:
\begin{align*}
\tilde{Z}^{6v}_N(\zeta_1,..,\zeta_N|z_1,..,z_N)= Z^{6v}_{N}(\zeta_1,..,\zeta_N|z_1,..,z_N)\prod _{i=1}^m \left(q^2-1\right)^{m-1} \zeta_i^{m-1} z_i^{m-1} ,
\end{align*} 
where we indicated explicitly the system size $N$ and put $6v$ in the superscript. 
$\tilde{Z}^{6v}_N$ has two recurrence relations 
that correspond to setting $\zeta_j=z_i$ and $\zeta_j=q^{-1}z_i$. For their derivation 
one can consult \cite{Izergin,Korepin} or see the explanation of similar recurrences 
in the case of the nineteen vertex model in Section \ref{sect4}.
The recurrence relations are:
\begin{align}\label{rec16v}
\tilde{Z}^{6v}_N(\zeta_1,..,\zeta_j=z_i,..,\zeta_N|z_1,..,z_N)=f_{i,j}^N\tilde{Z}^{6v}_{N-1}[\zeta_j,z_i],
\end{align} 
and
\begin{align}\label{rec26v}
\tilde{Z}^{6v}_N(\zeta_1,..,\zeta_j=q^{-1} z_i,..,\zeta_N|z_1,..,z_N)=g_{i,j}^N\tilde{Z}^{6v}_{N-1}[\zeta_j,z_i],
\end{align}
where the square brackets indicate which variables are absent from the initial list of 
variables on the left hand side.
The corresponding factors in the two recurrences are
\begin{align}
f_{i,j}^N=\prod_{1\leq k\neq i \leq N}(q^2 z_i^2-z_k^2)
\prod_{1\leq k\neq j \leq N}(q^2 \zeta_k^2-z_i^2), \\
g_{i,j}^N=\prod_{1\leq k\neq i \leq N}(z_i^2-q^2 z_k^2)
\prod_{1\leq k\neq j \leq N}(\zeta_k^2-z_i^2).
\end{align}
$\tilde{Z}^{6v}_N$ is expressed in terms of $\tilde{Z}^{6v}_{N-1}$'s as:
\begin{align}
&\tilde{Z}^{6v}_N(\zeta_1,..,\zeta_N|z_1,..,z_N)=\sum_{k=1}^{N}\tilde{Z}^{6v}_{N-1}[\zeta_N,z_k]
\prod_{i=1,i\neq k}^N\frac{(\zeta_N^2-z_i^2)}{(z_k^2- z_i^2)}f_{i,N}^N,~~~\text{and} \\
&\tilde{Z}^{6v}_N(\zeta_1,..,\zeta_N|z_1,..,z_N)=\sum_{k=1}^{N}\tilde{Z}^{6v}_{N-1}[\zeta_N,z_k]
\prod_{i=1,i\neq k}^N\frac{(q^2\zeta_N^2-z_i^2)}{(z_k^2- z_i^2)}g_{i,N}^N.
\end{align}
The recurrence Eq. (\ref{rec16v}) was derived by Korepin and solved by Izergin and the solution is written as the following determinant:
\begin{align}
&\tilde{Z}^{6v}_N=\mathcal{N} \det_{1\leq i,j \leq N}(\frac{1}{(\zeta_i^2-z_j^2)(q^2\zeta_i^2-z_j^2)}), \\
&\mathcal{N}=\frac{\prod_{1\leq i,j\leq N}(\zeta_i^2-z_j^2)
(q^2\zeta_i^2-z_j^2)}{\prod_{1\leq i<j\leq N}(\zeta_i^2-\zeta_j^2)
(z_j^2-z_i^2)}.\nonumber
\end{align}
$\tilde{Z}^{6v}_N$ is a homogenous polynomial in $z_1^2,..,z_N^2,\zeta_1^2,..,\zeta_N^2$ of degree $N(N-1)$
 and it satisfies the required recurrence relations together with the initial condition $\tilde{Z}^{6v}_1=1$.

\section{Nineteen-vertex model with domain wall boundary}\label{sect3}
\begin{figure}[htb]
\centering
\includegraphics[width=0.8\textwidth]{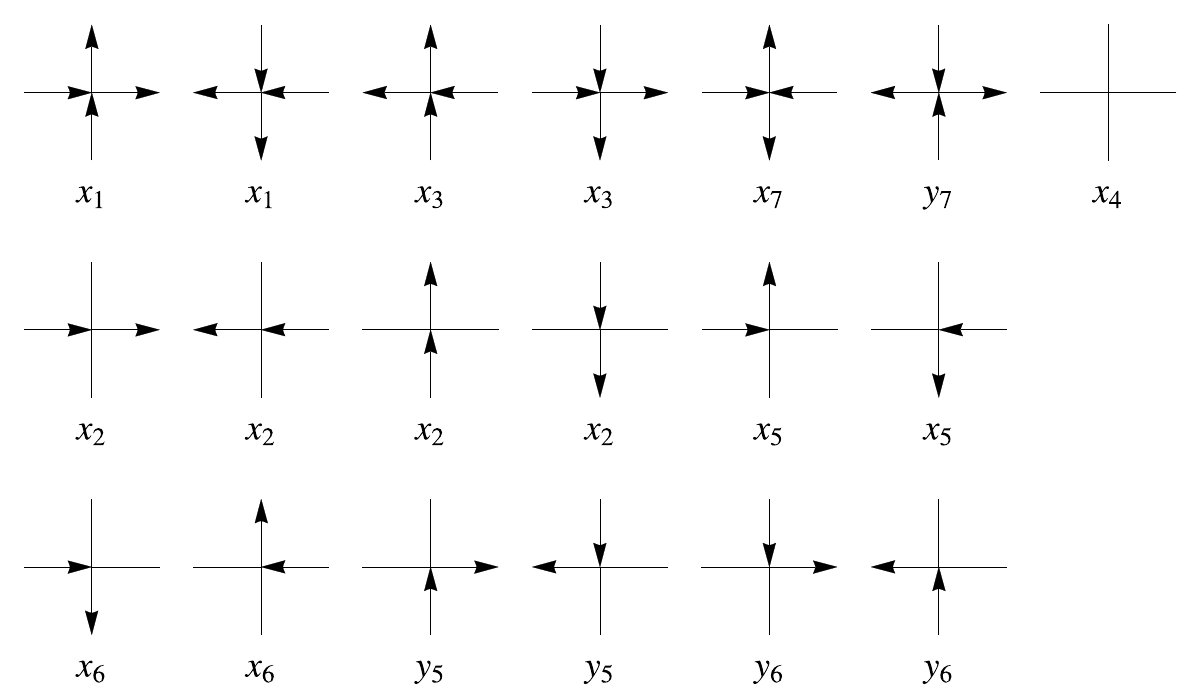}
\caption{The nineteen vertices and their weights.}
\label{vertices}
\end{figure}

Consider an inhomogeneous nineteen vertex model on a lattice. States of the 
model are defined through assigning one of the nineteen configurations to each vertex of the 
lattice. Each edge of the lattice can be in three states, denoted by arrows or an empty edge, 
in such a way that the total number of arrows pointing towards a vertex has to be equal 
to the total number of arrows pointing outwards. This restriction defines the nineteen possible 
configurations at each vertex Fig. \ref{vertices}.

The weights of the nineteen vertices are encoded in the $R$-matrix
\begin{align}
\left(
\begin{array}{ccccccccc}
 x_1(\zeta ) & 0 & 0 & 0 & 0 & 0 & 0 & 0 & 0 \\
 0 & x_2(\zeta ) & 0 & x_5(\zeta ) & 0 & 0 & 0 & 0 & 0 \\
 0 & 0 & x_3(\zeta ) & 0 & x_6(\zeta ) & 0 & x_7(\zeta ) & 0 & 0 \\
 0 & y_5(\zeta ) & 0 & x_2(\zeta ) & 0 & 0 & 0 & 0 & 0 \\
 0 & 0 & y_6(\zeta ) & 0 & x_4(\zeta ) & 0 & x_6(\zeta ) & 0 & 0 \\
 0 & 0 & 0 & 0 & 0 & x_2(\zeta ) & 0 & x_5(\zeta ) & 0 \\
 0 & 0 & y_7(\zeta ) & 0 & y_6(\zeta ) & 0 & x_3(\zeta ) & 0 & 0 \\
 0 & 0 & 0 & 0 & 0 & y_5(\zeta ) & 0 & x_2(\zeta ) & 0 \\
 0 & 0 & 0 & 0 & 0 & 0 & 0 & 0 & x_1(\zeta )\\
\end{array}
\right).
\end{align}
For the model under consideration they are defined as follows:
\begin{align}\label{weights19}
&x_1(\zeta )=\left(\zeta  q^2-1\right) \left(\zeta  q^3+1\right),\nonumber\\
&x_2(\zeta )=q \left(\zeta -1\right) \left(\zeta  q^3+1\right),\nonumber \\
&x_3(\zeta )=q^2\left(\zeta -1\right)   \left(\zeta q+1\right),\nonumber \\
&x_4(\zeta )=\zeta ^2 q^4+\zeta  (q-1) \left(q^4-q^2+1\right)-q,\nonumber \\
&x_5(\zeta )=\sqrt{\zeta} 
   \left(q^2-1\right) \left(\zeta q^3+1\right),\nonumber \\
&x_6(\zeta )=-\sqrt{q \zeta} 
   \left(\zeta-1\right)  \left(q^2-1\right),\nonumber \\
&x_7(\zeta )=\zeta
   \left(q^2-1\right) \left(\zeta q^3+\left(\zeta -1\right) q+1\right),\nonumber \\
&y_5(\zeta
   )=\sqrt{\zeta}  \left(q^2-1\right) \left(\zeta q^3+1\right),\nonumber \\ 
&y_6(\zeta )=q^{2}\sqrt{q\zeta}  \left(\zeta
   -1\right)  \left(q^2-1\right),\nonumber \\
&y_7(\zeta )=\left(q^2-1\right) \left(\zeta 
   q^3-\left(\zeta-1\right) q^2+1\right).
\end{align}
The graphical notation is completely analogous to the case of the six vertex model Fig. (\ref{fRc}). Below we will work with the $R$-matrix that depends on two parameters $R(z,\zeta)=z^2 R(\zeta/z)$, where $\zeta$ is the horizontal spectral parameter and $z$ is vertical.   
This $R$-matrix defines the integrable vertex model associated to the quantum group 
$U_q(A_2^{(2)})$ (see the derivation in \cite{KTuniq}). This model was discovered by Izergin and Korepin 
and thus called the Izergin--Korepin nineteen-vertex model. The corresponding $\check{R}$-matrix
satisfies the Yang--Baxter equation which is the same as before (\ref{YBm}).

We are interested in counting configurations of the following object. Consider the square 
lattice of size $N$ filled in with the above nineteen vertices in such a way that horizontal 
boundary arrows are pointing in to the lattice, while the vertical ones pointing outside the 
lattice. These boundary conditions are called, as before, the domain wall boundary 
Fig. \ref{DWPF}. 
The partition function of this object is the sum of all possible configurations with weights 
defined in Eq. (\ref{weights19}).

\begin{align}
Z_N=\sum_{\varepsilon\in \text{states}}\prod_{1\leq i,j\leq N}w^{(\varepsilon)}_{i,j},
\end{align}
where $w^{(\varepsilon)}_{i,j}$ is the weight of the vertex sitting at a position ${i,j}$ of 
a configuration $\varepsilon$.
This partition function is a symmetric polynomial in both horizontal $\zeta_i$ and vertical 
$z_i$ rapidities. The fact that it is a polynomial comes from the observation that each vertex 
that has a $\sqrt{\zeta}$ appears necessarily with another vertex that has a $\sqrt{\zeta}$. 
These weights are: $x_5,~x_6,~y_5$ and $y_6$ and they correspond to the vertices which have a 
``turning'' of an empty line. Clearly, the number of such turnings must be even in any DWPF 
configuration. The fact that $Z_N$ is symmetric can be proved as in the case of the six 
vertex model 
by attaching the $R$-matrix to two horizontal external lines in Fig. \ref{DWPF} or two vertical 
external lines and repetitive application of the Yang--Baxter equation. 
Hence the partition function $Z_N(\zeta_1,..,\zeta_N,z_1,..,z_N)$ is a symmetric polynomial 
in $z_i$'s and $\zeta_j$'s with coefficients being polynomials in $q$ with integer 
coefficients. 

\section{Recurrence relation}\label{sect4}
Before proceeding let us redefine $Z_N$ since it contains a number of trivial factors which we do not want to carry around. 
The variable $\zeta_i$ enters the partition function through the weights corresponding to the vertices located on the $i$-th horizontal line (counting from the top). The top degree in $\zeta_i$ of the weights $x_5,~x_6,~y_5$ and $y_6$ is equal to $3/2$ while for all other weights the top degree equals to $2$. Hence the top degree of $Z_N$ in $\zeta_i$ must be equal to $2N$ since there is always a configuration (e.g. six vertex configuration) which does not contain any of the vertices $x_5,~x_6,~y_5$ and $y_6$ in the $i$-th row. However, each configuration contains the trivial proportionality factor $\zeta_i$ which means that the top degree of the nontrivial part of the partition function is $2N-1$. Indeed, due to the boundary conditions each horizontal line must contain only an even number of the weights $x_5,~x_6,~y_5$ and $y_6$ including zero number of these weights. Each of these weights corresponding to the $i$-th line is proportional to $\zeta_i^{1/2}$, hence the weight of the configuration is divisible by $\zeta_i$. The configurations on the $i$-th horizontal line containing none of those weights necessarily contains the weight $x_7$ due to the boundary conditions. The weight $x_7(\zeta_i)\propto \zeta_i$. We conclude that the partition function has the factor $\zeta_1 .. \zeta_N$. It is convenient to consider $\tilde{Z}_N$
\begin{align*}
Z_N(\zeta_1..,\zeta_N;z_1,..,z_N)=(q^2 - 1)^N \prod_{i=1}^N \zeta_i \tilde{Z}_N (\zeta_1..,\zeta_N;z_1,..,z_N),
\end{align*} 
where the multiplication by $(q^2 - 1)^N$ accounts for another unimportant common factor. To define uniquely this partition function as a polynomial in $\zeta_i$ we need to know its values in $2N$ points which are provided by $2N$ recurrence relations derived below. 

The partition function $\tilde{Z}_N$ satisfies two recurrence relations in size with the initial 
condition $\tilde{Z}_0=1$\footnote{We extend the definition of the partition function $\tilde{Z}_N$ to the value $N=0$ and put $\tilde{Z}_0=1$ which is consistent with the recurrence relations derived in this section.}. They lead to: 
\begin{align}
\tilde{Z}_N(\zeta_1,..,\zeta_N|z_1,..,z_N)=\sum_{i=1}^{N} \kappa_i (\zeta_1,..,\zeta_N|z_1,..,z_N) 
\tilde{Z}_{N-1}(\zeta_1,..,\zeta_{N-1}|..,\hat{z_i},..), \label{recrel}
\end{align}
with some appropriate functions $\kappa_i$.

\begin{figure}[htb]
\centering
\includegraphics[width=0.3\textwidth]{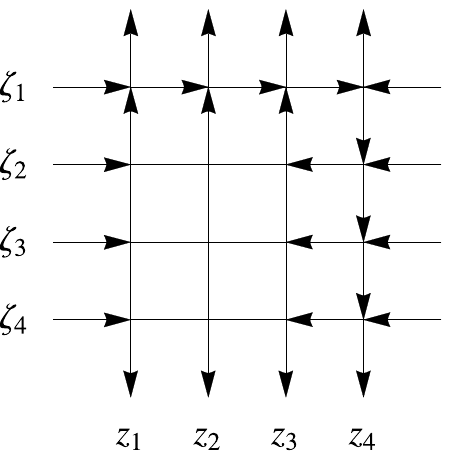}
\caption{The recurrence relation under substitution $\zeta_1=z_4$ for a $4\times 4$ lattice. 
Since the first row is frozen and the last column is frozen we obtain simple factors of 
$x_1$-weights, while the remaining configuration has the domain wall boundary conditions and 
corresponds to $Z_3$.}
\label{recfig}
\end{figure}

By inspecting the vanishing 
properties of the weights of the $R$-matrix we notice that there are two recurrence 
relations in size. When we set $\zeta_j$ to $z_i$ in $\tilde{Z}_n$ we get:
\begin{align}\label{rec1}
\tilde{Z}_N(\zeta_1,..,\zeta_j=z_i,..,\zeta_N|z_1,..,z_N)=F^N_{i,j}\tilde{Z}_{N-1}[\zeta_j,z_i],
\end{align}
This recurrence have a graphical interpretation shown 
in Fig. \ref{recfig}. Indeed, if we look at the north east corner (position $(1,N)$ on the lattice) 
of the domain, the boundary condition allows only for three vertices. 
These are vertices with the weights $x_3,~x_7$ and $x_6$. After 
setting $\zeta_1=z_N$, $x_3$ and $x_6$ vanish, so we are left with 
the vertex $x_7$. This vertex has a down arrow on its vertical lower edge and 
a right arrow on its left edge, hence 
due to the boundary condition at position $(2,N)$ we are forced to put the vertex corresponding 
to the weight $x_1$ and at position $(1,N-1)$ the other vertex with the weight $x_1$. 
In fact, all remaining vertices in the $N$-th column are frozen, as well as all the remaining 
vertices of the first row. These vertices contribute with the products of 
$x_1$-weights:
\begin{align}\label{factors}
\prod_{1\leq i\leq N-1}x_1(\zeta_1/z_i)\prod_{2\leq i\leq N}x_1(\zeta_i/z_N)|_{\zeta_1=z_N}.
\end{align}
A different recurrence appears when we set 
$\zeta_j$ to $-q^{-3}z_i$ in $\tilde{Z}_N$:
\begin{align}\label{recG}
\tilde{Z}_N(\zeta_1,..,\zeta_j=-q^{-3}z_i,..,\zeta_N|z_1,..,z_N)=G^N_{i,j}\tilde{Z}_{N-1}[\zeta_j,z_i],
\end{align}

The graphical explanation of this recurrence is similar to the previous recurrence. One 
must consider the top left corner of our domain and observe that only one vertex does not vanish 
under the substitution $\zeta_1=-q^{-3}z_1$. The first row and the first column freeze, while 
the rest returns the domain wall boundary condition for the domain of size $N-1\times N-1$.  

The $F$ and the $G$ are given by:
\begin{align}
F_{i,j}^N=(q^3+1)z_i\prod_{1\leq k\neq i \leq N}(q^2 z_i-z_k)(q^3 z_i+z_k)
\prod_{1\leq k\neq j \leq N}(q^2 \zeta_k-z_i)(q^3 \zeta_k+z_i), \label{Ff} \\
G_{i,j}^N=-q^{-N-1}(q^3+1)z_i\prod_{1\leq k\neq i \leq N}(z_i-q^2 z_k)(z_i+q^3 z_k)
\prod_{1\leq k\neq j \leq N}(\zeta_k-z_i)(q \zeta_k+z_i).\label{Gf}
\end{align}
If we know $\tilde{Z}_{N-1}$ these two recurrence relations allow us to determine $\tilde{Z}_N$. We can consider  
$\tilde{Z}_N$ as a polynomial in $\zeta_N$ of degree $2N-1$ with $2N$ coefficients. Since we know the values of $\tilde{Z}_N$ at $N$ points $\zeta_N=z_i$ (Eq. (\ref{rec1}) with $j=N$) and at another $N$ points $\zeta_N=-q^{-3}z_i$ (Eq. (\ref{recG}) with $j=N$), therefore we can determine all the coefficients of $\tilde{Z}_N$ in its 
expansion in $\zeta_N$. Using the Lagrange polynomial we can write $\tilde{Z}_N$ as a sum of $\tilde{Z}_{N-1}$'s 
as follows:
\begin{align}
&&\tilde{Z}_N(\zeta_1,..,\zeta_N|z_1,..,z_N)=\sum_{k=1}^{N}\tilde{Z}_{N-1}[\zeta_N,z_k]
\frac{\prod_{i=1}^N(\zeta_N-z_i)(\zeta_N+q^{-3} z_i)}{\prod_{i=1,i\neq k}^N(z_k- z_i)}\times
\nonumber \\ 
&&\bigg{(} \frac{F_{k,N}^{N}}
{(\zeta_N-z_k)\prod_{i=1}^N(z_k+q^{-3} z_i)}
 -  \frac{q^{3(N-1)}G_{k,N}^{N}}
{(\zeta_N+q^{-3} z_k)\prod_{i=1}^N(q^{-3}z_k+z_i)
}\bigg{)}.
\end{align}
This is of course a polynomial because the denominators are canceled by the common 
prefactor and by the $F$ and $G$ respectively including the factors of $q^3-1$ in the 
definitions Eq. (\ref{Ff}) and Eq. (\ref{Gf}). Using this we write
\begin{align}
&&\tilde{Z}_N(\zeta_1,..,\zeta_N|z_1,..,z_N)=q^{3(N-1)}\sum_{k=1}^{N}\tilde{Z}_{N-1}[\zeta_N,z_k]
\prod_{i=1,i\neq k}^N\frac{(\zeta_N-z_i)(\zeta_N+q^{-3} z_i)}{(z_k- z_i)}\times
\nonumber \\ 
&&\bigg{(}
(q^{3}\zeta_N+z_k)\prod_{i\neq k}(q^2 z_k - z_i)
\prod_{1\leq i\leq N-1}(-z_k+q^2\zeta_i)(z_k+q^3\zeta_i) + \nonumber \\ 
&&q^{2N-1}(\zeta_N-z_k)\prod_{i\neq k}(z_k -q^2 z_i)
\prod_{1\leq i\leq N-1}(-z_k+\zeta_i)(z_k+q\zeta_i)
\bigg{)}.
\label{rec2}
\end{align}
Possibly there is a way to 
write $\tilde{Z}_N$ for generic $q$ as a single determinant, for now this remains an open question. 
In the next section we show how to solve the recurrence relation for $\tilde{Z}_N$ when $q=e^{i\pi/3}$.

\section{Solution for a cubic root of unity}\label{sect5}
In this section we will assume $q=e^{i\pi/3}$. The recurrence relation Eq. (\ref{rec2}) simplifies in 
this case. Upon setting $q=e^{i\pi/3}$ (which implies $q^3 =-1$ and $q+q^{-1}=1$) the weights of the $R$-matrix (\ref{weights19}) have a common factor of $\zeta-1$ hence $Z_N$ factors out the product:
\begin{align}
\prod_{1\leq i,j \leq N}(z_i-\zeta_j).
\end{align}
We define  $\bar{Z}_N$ which has a more suitable normalization 
\begin{align*}
Z_N(\zeta_1..,\zeta_N;z_1,..,z_N)=-q^2 \prod_{1\leq i,j \leq N}(z_i-\zeta_j) \bar{Z}_N (\zeta_1..,\zeta_N;z_1,..,z_N),
\end{align*} 
The initial condition becomes $\bar{Z}_1=1$ and out of the two recurrence points only one 
remains. Setting $q=e^{i\pi/3}$ in eq. (\ref{rec2}) we find
\begin{align*}
&\tilde{Z}_N(\zeta_1,..,\zeta_N|z_1,..,z_N)=(-1)^{N-1}\sum_{k=1}^{N}\tilde{Z}_{N-1}[\zeta_N,z_k]
\prod_{i=1,i\neq k}^N\frac{(\zeta_N-z_i)^2}{(z_k- z_i)}\times
\prod_{i=1}^N (z_k-\zeta_i)\times\nonumber \\ 
&\bigg{(}
\prod_{i\neq k}(q^2 z_k - z_i)
\prod_{1\leq i\leq N-1}(-z_k+q^2\zeta_i) + (-1)^N q^{2N-1}\prod_{i\neq k}(z_k -q^2 z_i)
\prod_{1\leq i\leq N-1}(z_k+q\zeta_i)
\bigg{)}.
\end{align*}
Rewriting this in terms of $\bar{Z}_N$, after cancellations we get 
\begin{align*}
&\bar{Z}_N(\zeta_1,..,\zeta_N|z_1,..,z_N)=\sum_{k=1}^{N}\bar{Z}_{N-1}[\zeta_N,z_k]
\prod_{i=1,i\neq k}^N\frac{(\zeta_N-z_i)}{(z_k- z_i)}\nonumber \\ 
&\bigg{(}
\prod_{i\neq k}(q^2 z_k - z_i)
\prod_{1\leq i\leq N-1}(-z_k+q^2\zeta_i) + (-1)^N q^{2N-1}\prod_{i\neq k}(z_k -q^2 z_i)
\prod_{1\leq i\leq N-1}(z_k+q\zeta_i)
\bigg{)}.
\end{align*}
Set here, say $\zeta_N=z_{N}$, and put $z_N=x$. In the right hand side only one term survives in the summation over $k$
\begin{align*}
&\bar{Z}_N(\zeta_1,..,\zeta_{N-1},x|z_1,..,z_{N-1},x)=\bar{Z}_{N-1}(\zeta_1,..,\zeta_{N-1}|z_1,..,z_{N-1})
\nonumber \\ 
&\bigg{(}
\prod_{1\leq i\leq N-1}(q^2 x - z_i)(-x+q^2\zeta_i) + (-1)^N q^{2N-1}
\prod_{1\leq i\leq N-1}(x -q^2 z_i)(x+q\zeta_i)\bigg{)} =\nonumber \\
&\bar{Z}_{N-1}(\zeta_1,..,\zeta_{N-1}|z_1,..,z_{N-1})(-1)^N q^{2N-1}
\nonumber \\ 
&\bigg{(}
-q^{-1}\prod_{1\leq i\leq N-1}(z_i+ x/q )(q x+\zeta_i) + 
\prod_{1\leq i\leq N-1}(q x+  z_i)(x/q+\zeta_i)
\bigg{)} =\nonumber \\
&\bar{Z}_{N-1}(\zeta_1,..,\zeta_{N-1}|z_1,..,z_{N-1})(-1)^N q^{2N}
\nonumber \\ 
&\bigg{(}
q\prod_{1\leq i\leq N-1}(z_i+ x/q )(q x+\zeta_i) + q^{-1}
\prod_{1\leq i\leq N-1}(q x+  z_i)(x/q+\zeta_i).
\bigg{)} 
\end{align*}
where we used the fact that  $q=e^{i\pi/3}$.
By the symmetry we can  interchange $\zeta_N$ with $\zeta_j$ therefore we find
\begin{align}
\bar{Z}_N(\zeta_1,..,\zeta_j=z_i,..,\zeta_N|z_1,..,z_N)=P_{i,j}\bar{Z}_{N-1}[\zeta_j,z_i] \label{recw}.
\end{align}
Let us focus on the polynomial $P_{i,j}$ and for convenience we specify $i=N$, $j=N$ and 
set $z_N=x$. The polynomial $P_{N,N}=P(x|\zeta_1,..,\zeta_{N-1},z_1,..,z_{N-1})$ is a symmetric 
polynomial in $\zeta_1,..,\zeta_{N-1}$ and separately in $z_1,..,z_{N-1}$
\begin{align}
& P(x|\zeta_1,..,\zeta_{N-1},z_1,..,z_{N-1})= 
\nonumber \\
& (-1)^{N}q^{2N} \left(q \prod _{i=1}^{N-1}  \left(\zeta _i+q x\right) \prod _{i=1}^{N-1}
   \left(z_i+x/q\right)+\frac{1}{q}\prod _{i=1}^{N-1} \left(\zeta _i+x/q \right)
   \prod _{i=1}^{N-1} \left(z_i+q x\right)\right).\label{Prec}
\end{align}
Note, up to the overall factor of $q^{2N}$, $P$ is invariant under $q\rightarrow 1/q$, which means 
it has to be a function of $q^i+q^{-i}$. When  $q^3 =-1$ the combinations $q^i+q^{-i}$ are integers, hence $P$ becomes a 
polynomial with purely integer coefficients. The same is therefore also true for $\bar{Z}_N$ 
itself. Let us consider now $P$ as the generating function for some symmetric polynomials:
\begin{align}
&P_N(x)=P(x|\zeta_1,..,\zeta_{N},z_1,..,z_{N})= (-1)^{N}q^{2N}\sum_{i=0}^{2N}x^{i}
\Delta_{2N-i,N}(\zeta_1,..,\zeta_N,z_1,..,z_N) \label{gen}.
\end{align}
We included here the factor of $q^{N}$ in order to make $\Delta_{i,N}$ $q$-independent.
The polynomials $\Delta_{i,N}$ are polynomials of $2N$ variables with the total degree $i$. 
If $i<0$ or $i>2N$ we set it equal to $0$, and also $\Delta_{0,N}=1$. Here is the example for 
$N=2$:
\begin{align}
&\Delta_{1,2}=2 \zeta _1+2 \zeta _2-z_1-z_2,\nonumber \\
&\Delta_{2,2}=\zeta _1 \zeta _2+\zeta _1 z_2+\zeta _2 z_2+\zeta_1 z_1+\zeta _2 z_1-2 z_1 z_2,
\nonumber\\
&\Delta_{3,2}=-\zeta _1 z_1 z_2+2 \zeta _2 \zeta _1 z_1+2
   \zeta _2 \zeta _1 z_2-\zeta _2 z_1 z_2,\nonumber \\
&\Delta_{4,2}=\zeta _1 \zeta _2 z_1 z_2.\nonumber
\end{align}
These symmetric functions have a few nice properties which we will discuss in the next section. 
The solution of the recurrence relation $(\ref{recw})$ is the main result of our paper, it reads: 
\begin{align}
\bar{Z}_N(\zeta_1,..,\zeta_{N},z_1,..,z_{N})=
\det_{1\leq i,j \leq N-1}\Delta_{3j-i,N}(\zeta_1,..,\zeta_{N},z_1,..,z_{N}) \label{det},
\end{align}
or equivalently
\begin{align*}
\bar{Z}_N =\det_{1\leq i,j \leq N-1} \sum_ {l = 0}^
   N (q^{-(3j-i) + 2 l + 1} + q^{3j-i - 2 l - 1}) E_l (z_ 1, .., z_N) E_{3j-i- l} (\zeta_ 1, .., \zeta_N),
\end{align*}
where $E_{i}(x_1,..,x_N)$ are the elementary symmetric polynomials defined by $E_{i}(x_1,..,x_N)=0$ if $i<0$ or $i>N$ and otherwise
\begin{align}
E_{i}(x_1,..,x_N)=\sum_{1\leq n_1<..<n_i\leq N}x_{n_1}x_{n_2}..x_{n_i}.
\end{align}
The proof of this result follows next.

\section{Proof}\label{sect6}
Let us list few properties of $\Delta_{i,N}$. First of all, looking at the definition of 
these polynomials (\ref{gen}) we can immediately express them through the elementary symmetric 
polynomials $E_{i}$. The elementary symmetric polynomials can be defined via their generating function 
 \begin{align}
&Q^z(x)=Q(x;z_1,..,z_n)=\prod_{i=1}^N (x-z_i),\label{Qesp1}\\
&Q^z(x)=\sum_{n=0}^N (-1)^{N-n} x^n E_{N-n}(z_1,..,z_N).\label{Qesp2}
\end{align}
We can rewrite (\ref{Prec}) using $Q^z(x)$ and $Q^{\zeta}(x)$ 
 \begin{align*}
P(x;\zeta_1,..,\zeta_N,z_1,..,z_N)=(-q^2)^N \left( q Q^{\zeta}(x q)Q^{z}(x/q) +q^{-1} Q^{\zeta}(x/q)Q^{z}(x q) \right).
\end{align*}
Substituting Eq. (\ref{Qesp2}), expanding in $x$, collecting the coefficients and comparing them to (\ref{gen}) we get
\begin{align*}
\Delta_{2N-i,N}=\sum_{\substack{
            1\le n_1,n_2 \le N\\
            n_1+n_2=i}}^N (q^{1-n_1+n_2}+q^{-1+n_1-n_2})
E_{N-n_1}(\zeta_1,..,\zeta_N)E_{N-n_2}(z_1,..,z_N).
\end{align*}
The summation can be simplified and we obtain
\begin{align}
\Delta_{s,N} = \sum_ {l = 0}^
   N (q^{-s + 2 l + 1} + q^{s - 2 l - 1}) E_l (z_ 1, .., z_N) E_{s - l} (\zeta_ 1, .., \zeta_N),\label{delta}
\end{align}

Note that Eq. (\ref{delta}) is valid for generic values of $q$. When $q=1$, $\Delta_i$
becomes the elementary symmetric polynomials of the union of $z$'s and $\zeta$'s 
times a factor of two. So, it can be considered as a type of $q$-deformation of the elementary 
symmetric polynomials.  

Next, we  set $\zeta_N=z_N$ (without loss of generality). From the definition of $P_N$ we 
see that it produces back $P_{N-1}$: 
\begin{align}
P(x|\zeta_1,..,\zeta_{N},z_1,..,z_N)|_{\zeta_N=z_N}&=-(z_N q+x)(z_N+q x)
P(x|\zeta_1,..,\zeta_{N-1},z_1,..,z_{N-1})\nonumber \\
&=-q(z_N^2+x z_N+ x^2) P(x|\zeta_1,..,\zeta_{N-1},z_1,..,z_{N-1})\label{recP},
\end{align}
where in the second line we took into account that $q+q^{-1}=1$. Looking at 
Eq. (\ref{recP}) we can relate the set of $\Delta_{i,N}$'s in which 
$\zeta_N=z_N$ to the set of $\Delta_{j,N-1}$'s:
\begin{align}
&\Delta_{i,N}(\zeta_1,..,\zeta_N=z_N,z_1,..,z_N)=
\Delta_{i,N-1}(\zeta_1,..,\zeta_{N-1},z_1,..,z_{N-1})\nonumber \\
&+z_N \Delta_{i-1,N-1}(\zeta_1,..,\zeta_{N-1},z_1,..,z_{N-1})+
z_N^2 \Delta_{i-2,N-1}(\zeta_1,..,\zeta_{N-1},z_1,..,z_{N-1}) \label{recd}
\end{align}
Using this equation and a certain row-column manipulation in the matrix $\Delta_{3j-i,N}$ 
we are going to show that the determinant (\ref{det}) satisfies the recurrence (\ref{recw}). 

Set $\zeta_N=z_N$ and substitute Eq. (\ref{recd}) in every entry of the matrix 
in Eq. (\ref{det}). Starting from the first row 
subtract from each row $i$ row $i+1$ multiplied by $z_N$. Next, subtract from each column 
$j$ column $j+1$ multiplied by $z_N^{3(N-1-j)}$ starting from the $j=(N-2)$-th column. In the 
resulting matrix all elements of the first column become zero except from the bottom element. The 
bottom element in the first column takes the form of Eq. (\ref{gen}), while the rest of the 
matrix equals to $\Delta_{3j-i,N}$ of size $N-1$, and the last row is unimportant upon 
taking the determinant. The row-column manipulation above corresponds to the following series 
of equations. Upon application of the recurrence relation each entry becomes:
\begin{align}\label{subd}
\Delta_{3j-i,N-1}+z_N \Delta_{3j-i-1,N-1}+z_N^2 \Delta_{3j-i-2,N-1} 
\end{align}
After the first row manipulation the last row remains as before: 
\begin{align}
\Delta_{3j-N+1,N-1}+z_N \Delta_{3j-N,N-1}+z_N^2 \Delta_{3j-N-1,N-1}, 
\end{align}
the rest part of the matrix becomes:
\begin{align}
\Delta_{3j-i,N-1}-z_N^3 \Delta_{3j-i-3,N-1}. 
\end{align}
We notice that in the last column the first of these two terms vanish $\Delta_{3(N-1)-i,N-1}$ 
for all $i<N-1$, while in the first column the second term 
vanish. Next, we use the last column to eliminate 
the unwanted terms in other entries of the matrix (except from the last row). After this,  
the first column except from its last element will vanish, while the last element will be:
\begin{align}
&\sum_{j=1}^{N-1}z_{N}^{3(N-1-j)}(\Delta_{3j-N+1,N-1}+z_N \Delta_{3j-N,N-1}+z_N^2 
\Delta_{3j-N-1,N-1})\nonumber\\
&=q^{2(N-1)} P(z_N|\zeta_{1},..,\zeta_{N-1},z_1,..,z_{N-1})
\end{align}
This completes the proof. We can alternatively view this row column manipulation as acting on 
the left and on the right of Eq. (\ref{subd}) with certain matrices with unit determinant. 
Let us call the expression in Eq. (\ref{subd}) $\tilde{\Delta}_{3j-k,N-1}$, and define two matrices:
\begin{align}
A=\left(
\begin{array}{ccccc}
 1 & -z & 0 & \dots & 0 \\
 0 & 1 & -z & \dots & 0 \\
 \dots & \dots & \dots & \dots & \dots \\
 0 & 0 & 0 & \dots & -z \\
 0 & 0 & 0 & \dots & 1 \\
\end{array}
\right)
\end{align}
and 
\begin{align}
B=\left(
\begin{array}{ccccc}
 1 & z^3 & z^6 & \dots & z^{3(N-1)} \\
 0 & 1 & z^3 & \dots & z^{3(N-2)} \\
 \dots & \dots & \dots & \dots & \dots \\
 0 & 0 & 0 & \dots & z^3 \\
 0 & 0 & 0 & \dots & 1 \\
\end{array}
\right)
\end{align}
We have:
\begin{align}
\det_{1 \leq k,l \leq N-1}A_{i,j} \tilde{\Delta}_{3j-k,N-1}B_{k,l}=
q^{2(N-1)} P_{N-1}(z_N)
\det_{1 \leq k,l \leq N-2} \Delta_{3j-k,N-1}.
\end{align}

\section{Conclusion}
As we mentioned in the introduction, our study is motivated by the six vertex model. Hence, 
it is natural to look at other related objects which were computed for the six vertex model. 
Since the nineteen vertex model seems to have a more complicated structure, one probably 
should not expect to obtain nice answers as in the six vertex case. 
As we have observed, however, when $q$ is a root of unity the nineteen vertex model 
becomes ``\textit{computable}". 
 
Here we considered the domain wall boundary conditions for the nineteen vertex model of 
Izergin and Korepin. An interesting extension of our computation would be to consider 
other boundary conditions, i.e. to use reflection matrices on one or two sides of the $N\times N$ 
domain. In the case of the six vertex model the corresponding partition functions are known 
to be determinants or Pfaffians (see \cite{Tsuchiya} and also \cite{Kuperberg2}). 
One would need to find first the recurrence relation for the partition function and then after
setting $q=e^{i\pi/3}$ it should be possible to obtain a determinantal expression. 
We note here that similar determinants appear in the study of the related loop model 
exactly when $q=e^{i\pi/3}$. The loop model related to the IK model 
is called the dilute Temperley--Lieb (dTL($n$)) loop model \cite{Nienhuis1,Nienhuis2}. This model has a parameter: the 
weight $n$ of a loop. When $q=e^{i\pi/3}$ this corresponds to $n=1$ and the corresponding loop model is 
related to interesting statistical models like critical percolation, for example.  
In \cite{GN} it was shown that the partition function of the dTL($1$) model 
satisfies a similar recurrence as in Eq. (\ref{recw}) and Eq. (\ref{Prec}), and has a solution 
similar to Eq. (\ref{det}). Recurrence relations of the form Eq. (\ref{recw}) also appear in the study of the 3-state Potts model \cite{KSm}.

\begin{figure}[htb]
\centering
\includegraphics[width=0.9\textwidth]{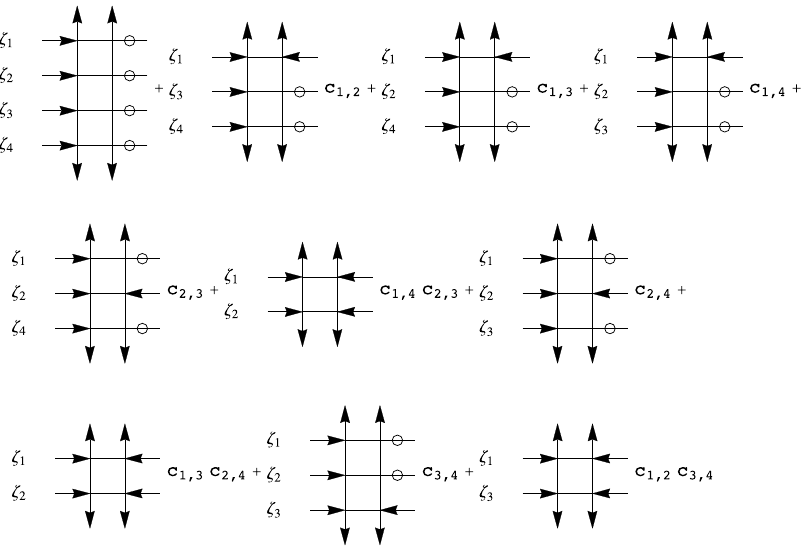}
\caption{This is a graphical representation of the highest spin off-shell eigenvector of the IK transfer matrix. The coefficients $c_{i,j}$ are certain 
functions of the weights in Eq. (\ref{weights19}). The circles appearing on the right boundaries  
signify that the corresponding edges are in the empty state.}
\label{hev}
\end{figure}

In the context of the algebraic Bethe Ansatz the domain wall partition function for the six 
vertex model represents the highest spin eigenvector of the corresponding transfer matrix 
with periodic boundary conditions. The parameters $\zeta_i$ become the Bethe roots. This object 
is essential in the study of correlation functions of the corresponding model. One may 
similarly look at 
the highest spin eigenvector of the transfer matrix for the IK model given 
by the corresponding algebraic Bethe Ansatz (see \cite{Tarasov} and also \cite{LS}). However, for the nineteen 
vertex model the eigenvectors of the transfer matrix are much more complicated than in the 
case of the six vertex model. For example, to compute highest spin eigenvector 
we need to consider the nineteen vertex model with many different boundary conditions 
on rectangular domains. 
The expression for this eigenvector for $N=4$ pictorially is shown in Fig. \ref{hev}.
In general, the expression for this eigenvector looks very complicated. For a root of minus one $q^3=-1$ we know few terms here, i.e. those corresponding to the domain wall boundaries. The 
other terms should not be expected to have a nice closed form since they, in general, are not 
symmetric in $\zeta$'s nor in $z$'s. The eigenvector as a whole is symmetric in $\zeta$'s and 
$z$'s. One then needs to find a recurrence relation for it and then, if lucky, it will 
be possible to find its closed form solution at $q=e^{i\pi/3}$. The knowledge of this will be helpful 
in understanding of the other eigenvectors. In particular, we could look at the zero 
spin eigenvectors at $q=e^{i\pi/3}$ (i.e. when lower boundary has equal number of up arrows 
and down arrows). One such eigenvector was computed in the loop basis \cite{GN2} by 
means of the quantum Knizhnik--Zamolodchikov equations.

Let us look carefully at the determinant expression in (\ref{det})
\begin{align*}
\bar{Z}_N =\det_{1\leq i,j \leq N-1} \sum_ {l = 0}^
   N (q^{-(3j-i) + 2 l + 1} + q^{3j-i - 2 l - 1}) E_l (z_ 1, .., z_N) E_{3j-i- l} (\zeta_ 1, .., \zeta_N).
\end{align*}
This is very close to the Lascoux determinant \cite{Lasc} for the six vertex domain wall partition function. In particular, it can be rewritten as a determinant of a product of two rectangular matrices which can be computed applying the Cauchy--Binet formula. In this way it was shown in \cite{FWZ} that the six vertex IK determinant is a Kadomtsev--Petviashvili (KP) tau-function. Therefore, it would be natural to ask if the determinant (\ref{det}) is a tau-function.

Finally, regarding the generic $q$ expression for $Z_N$ partition function one could try 
to look for its expansion in terms of symmetric polynomials. For example, it is known 
that $Z_{6v}$ expands naturally in the Hall--Littlewood polynomials \cite{Warnaar,BW,BWZ}.

\section*{Acknowledgements}
The author is grateful to Michael Wheeler and Paul Zinn-Justin for inspiring discussions and to Nikolai Kitanine for pointing to various inaccuracies in the initial version of the text. The 
work is supported by the ERC grant 278124 ``Loop models, Integrability and Combinatorics''. 

\small{\bibliographystyle{plain}}

\end{document}